\documentclass[USletter, 10 pt, conference]{ieeeconf}

\IEEEoverridecommandlockouts                              
\usepackage{amsfonts}
\usepackage{amsmath}
\usepackage{amssymb}
\usepackage{graphicx}%
\usepackage[latin1]{inputenc}
\usepackage{epstopdf}
\usepackage{color}
\usepackage{multicol}
\usepackage{subcaption}
\usepackage{float}
\usepackage{csquotes}
\usepackage[dvipsnames]{xcolor}
\usepackage{tikz}

\newtheorem{theorem}{Theorem}[section]

\newcommand{\ben}{\begin{enumerate}}
	\newcommand{\een}{\end{enumerate}}
\newtheorem{thm}{Theorem}
\newtheorem{cor}{Corollary}
\newtheorem{lem}{Lemma}

\newtheorem{dfn}{Definition}

\newtheorem{rmq}{Remark}

\newtheorem{expl}[thm]{Example}
\newcommand{\R}{\mathbb{R}}

\newcommand{\blem}{\begin{lem}}
	\newcommand{\elem}{\end{lem}}
\newcommand{\bdfn}{\begin{dfn}}
	\newcommand{\edfn}{\end{dfn}}
\newcommand{\bcor}{\begin{cor}}
	\newcommand{\ecor}{\end{cor}}
\newcommand{\bthm}{\begin{thm}}
	\newcommand{\ethm}{\end{thm}}
\newcommand{\bex}{\begin{expl}}
	\newcommand{\eex}{\end{expl}}
\newcommand{\brmq}{\begin{rmq}}
	\newcommand{\ermq}{\end{rmq}}
\newcommand{\bitem}{\begin{itemize}}
	\newcommand{\eitem}{\end{itemize}}

\title{\LARGE \bf Stabilization of an unstable reaction-diffusion PDE with input delay despite state and input quantization}
\author{	Florent Koudohode and Nikolaos Bekiaris-Liberis\\
	\thanks{ 
			Funded by the European Union (ERC, C-NORA, 101088147). Views and opinions expressed are however those of the authors only and do not necessarily reflect those of the European Union or the European Research Council Executive Agency. Neither the European Union nor the granting authority can be held responsible for them.	
		\newline{Florent Koudohode (\tt\scriptsize fkoudohode@tuc.gr)} {and Nikolaos Bekiaris-Liberis (\tt\scriptsize bekiaris-liberis@ece.tuc.gr)} are with  Department of Electrical and Computer Engineering, Technical University of Crete, 73100, Greece.}
		 }
\begin{document}
	
	\maketitle
	\thispagestyle{empty}
	\pagestyle{empty}
	
	\begin{abstract}
	We solve the global asymptotic stability problem of an unstable reaction-diffusion Partial Differential Equation (PDE) subject to input delay and state quantization developing a switched predictor-feedback law. To deal with the input delay, we reformulate the problem as an actuated transport PDE coupled with the original reaction-diffusion PDE. Then, we design a quantized predictor-based feedback mechanism that employs a dynamic switching strategy to adjust the quantization range and error over time. The stability of the closed-loop system is proven properly combining backstepping with a small-gain approach and input-to-state stability techniques, for deriving estimates on solutions, despite the quantization effect and the system's instability. We also extend this result to the input quantization case. 

	\end{abstract} 

\section{Introduction}
The reaction-diffusion PDE is widely employed across various fields, including biology \cite{britton1986reaction}, chemistry \cite{grzybowski2009chemistry}, and epidemiology \cite{jiang2020136} to model phenomena such as heat distribution, chemical reactions, and the spread of diseases. Due to its broad applicability, the reaction-diffusion PDE has been extensively studied, in particular, in terms of stabilization under digital implementation effects, see, for example, \cite{espitia2021event,karafyllis2018sampled,katz2022sampled,Lhachemi,prieur2018feedback,rathnayake2022sampled,selivanov2018boundary} and references therein. In digital implementations of feedback laws for reaction-diffusion PDEs, the resulting closed-loop systems may be subject to input delay and quantization. Consider, for example, the case of epidemics spreading described by a reaction diffusion PDE system \cite{jiang2020136}, which may be subject to delay, due to, e.g., testing or measures imposition lags \cite{andrew,young2019consequences} and quantization, due to, e.g., discrete quarantine measures imposition (as implementation of measures chosen from a continuum are practically infeasible). The presence of input delay and quantization may deteriorate performance or even destabilize the closed-loop system. 

For this reason, besides results on control of systems under quantization, see, for instance, \cite{brockett2000quantized,delchamps1990stabilizing,liu2015dynamic} that examine linear systems with quantized measurements and \cite{liberzon2003hybrid} that addresses general nonlinear systems under quantization, there are results dealing with the effect of quantization in delay systems. In particular, \cite{fridman2009control} focuses on linear time-delay systems with saturation, \cite{garcia2012model} investigates logarithmic quantizers within an event-based control framework, and \cite{di2020practical} considers nonlinear systems with state delay and quantization. Stabilization of various classes of PDEs under static quantization has also been addressed. In particular, \cite{espitia2017stabilization} and \cite{tanwani2016input} address first-order hyperbolic systems, whereas \cite{katz2022sampled} and \cite{selivanov2016predictor} focus on parabolic systems. Dynamic quantizers with adjustable ``zoom" parameter have been designed for first-order hyperbolic systems in \cite{bekiaris2020hybrid}, for a linear ODE (Ordinary Differential Equation)-transport PDE cascade in \cite{fkoudohode2024}, and for a general class of infinite-dimensional, discrete-time systems in \cite{wakaiki2024stabilization}. One of the key differences between static and dynamic quantizers lies in that the latter can guarantee global asymptotic stability, which requires however assuming that the quantizer's range/error can be dynamically adjusted.

In this paper, we develop a switched predictor-feedback law for simultaneous compensation of input delay and state/input quantization. The feedback law is essentially a quantized version of the nominal backstepping/predictor-feedback law from \cite{krstic2009control}, which is constructed by properly designing the adjustable parameter of the quantizer in a time-triggered manner, as it is performed in \cite{fkoudohode2024} for ODE systems with input delay. Even though the control design introduced is inspired by \cite{fkoudohode2024} (and \cite{krstic2009control}), the feedback law developed here is new. In particular, we properly define the quantizer's properties in an infinite-dimensional setting to account for the norm that the system under investigation is equipped with, as well as we properly choose the parameters of the adjustable parameter of the quantizer based on open- and closed-loop solutions' estimates that we derive for the class of systems considered. We establish global asymptotic stability in $L^2$ norm of the PDE state and in $L^{\infty}$ norm for the actuator state. The stability analysis relies on a combination of input-to-state stability and small-gain arguments, with constructive derivation of solutions' estimates.

Section~\ref{probFormulation} introduces the class of systems and quantizers considered, along with the switched predictor-feedback design. Section~\ref{stabStateQuantization} and Section~\ref{inputquantization} establish global asymptotic stability of the closed-loop system under state and input quantization, respectively. Section~\ref{conlusion}  provides concluding remarks and suggests potential future research directions. 

 {\em Notation:} We denote by $L^2(0,1)$ the equivalence class of Lebesgue measurable functions $f: [0,1] \rightarrow \R$ such that $\Vert f\Vert_2 =\left(\int_{0}^{1}  \vert f(x)\vert ^{2}dx\right)^{1/2} < \infty.$ 
 For a given function  $u \in L^{\infty}([0, D] ; \R)$ we define $\|u\|_{\infty}=\operatorname{ess\ sup}_{x \in[0, D]} |u(x)| $ where $\operatorname{ess\ sup},$ is the essential supremum. The state space $L^2([0, 1] ; \R) \times L^{\infty}([0, 1] ; \R) $ is induced with norm $\|(u,v)\|=\Vert u \Vert_2 + \Vert v\Vert_{\infty} .$ 
  We denote by $\mathcal{C}_{\rm rpw}(I;\R)$ the space  of right piecewise continuous functions $f: I\to \R$ (see \cite{KKnonlocal}). 
For a given $h \in \R$ we define its integer part as $\lfloor h\rfloor=\max \{k \in \mathbb{Z}: k \le h\}$.
\section{ Problem Formulation and Control Design}\label{probFormulation}
\subsection{ Backstepping Control of Reaction-Diffusion PDE With Input Delay } 

Let us consider the following  scalar reaction-diffusion PDE with  known  constant input delay $D>0$
\begin{eqnarray}\label{eq:sysparabolic0_delay}
	u_t(t,x) & =&   u_{xx}(t,x) + \lambda u(t,x), \\
	u(t,0)&=&0, \label{BC_parabolic_PDE_u00_delay} \\
	u(t,1)&=& U(t- D),  \label{BC_parabolic_PDE_u0_delay}
\end{eqnarray} 
where $\lambda>\pi^2$ such that the open-loop system \eqref{eq:sysparabolic0_delay}--\eqref{BC_parabolic_PDE_u0_delay} is unstable. We pose this delay problem as an actuated transport PDE (modeling the delay phenomenon) which cascades into the  boundary of the reaction-diffusion PDE,
\begin{eqnarray}
	u_t(t,x) & =&   u_{xx}(t,x) + \lambda u(t,x), \label{eq:sysparabolic1_cascade}\\
	u(t,0)&=&0, \label{eq:BC_parabolic1_cascade} \\
	u(t,1)&=& v(t,0),  \label{eq:BC_parabolic1_cascadeuv} \\
	v_{t}(t,x)&=& \tfrac{1}{D}v_{x}(t,x),\label{eq:hyperbolic_cascade}\\
	v(t,1)&=&U(t),\label{eq:BC_hyperbolic_cascade}
\end{eqnarray} 
where $(t,x) \in \mathbb{R}_{+} \times [0,1]$, $u(t,\cdot)$ and $v(t,\cdot)$ are respectively, the reaction-diffusion PDE and the transport PDE states at time $t$, with initial  conditions $ u(0,x)=u_{0}(x) $ and $v(0,x)=v_{0}(x)$, $x \in [0,1]$, and variable $U(t) $ is control input.

Consider the following backstepping transformations 
\begin{align}
	w(t,x)&=u(t,x)-\int_0^xk(x,y)u(t,y)dy, \label{eq:direct_transformation1} \\
\nonumber	z(t,x)&=v(t,x)-D\int_0^x g(x,y)v(t,y)dy \\
	&-\int_0^1 \gamma(x,y)u(t,y)dy,\label{eq:direct_transformation2}
\end{align}
$x\in [0,1]$, with $\gamma(x,y)$, $k(x,y)$, and $g(x,y)$ given by
\begin{align}
\nonumber	\gamma(x,y)&=2 \sum_{n=1}^{\infty} e^{D(\lambda-n^2\pi^2)x}\sin(n\pi y)\\
	&\times\int_0^1\sin(n\pi \zeta) k(1,\zeta)d\zeta,\label{eq:kernel_gamma}
\end{align}
with
\begin{align} \label{eq:kernel_heat_bessel_function} 
	k(x,y)=-\lambda y \tfrac{I_1\left(\sqrt{\lambda(x^2-y^2)}\right)}{\sqrt{\lambda( x^2 -y^2)}},
\end{align}  
on $\mathcal{T}:=\{(x,y):0\leq y\leq x\leq 1\}$, where $\mathrm{I}_1(\cdot)$ denotes the modified Bessel function of first kind. In addition 
\begin{equation}\label{eq:kernel_q}
	g(x,y)=-\gamma_y(x-y,1).
\end{equation}
System \eqref{eq:sysparabolic1_cascade}--\eqref{eq:BC_hyperbolic_cascade} is transformed into the following system:
\begin{eqnarray}
	w_t(t,x) & =&   w_{xx}(t,x), \label{eq:Target_sysparabolic1_cascade}\\
	w(t,0)&=&0,  \\
	w(t,1)&=& z(t,0),  \label{eq:Target_BC_parabolic1_cascade} \\
	z_{t}(t,x)&=& \tfrac{1}{D}z_{x}(t,x),\label{eq:Target_hyperbolic_cascade}\\
	z(t,1)&=&d(t),\label{eq:Target_BC_hyperbolic_cascade}
\end{eqnarray} with initial conditions    
\begin{align} 
	w_0(x)&=u_0(x)-\int_0^xk(x,y)u_0(y)dy, \label{eq:Initial_cond_target_w} \\
\nonumber	z_0(x)&=v_0(x)  -\int_0^1 \gamma(x,y)u_0(y)dy\\
	  &- D\int_0^x g(x,y)v_0(y)dy, \label{eq:Initial_cond_target_z}
\end{align} and the deviation $d(t)$ is defined by
\begin{equation}
	d(t)=:U(t)-U_{\rm nom}(t),
\end{equation}
where $U_{\rm nom}(t)$ is the nominal predictor-feedback law  \begin{equation}\label{nominalU}
	U_{\rm nom}(t)=\int_0^1\gamma (1,y)u(t,y)dy+D\int_0^{1}g(1,y)v(t,y)dy.
\end{equation}
The  inverse transformation is given by 
\begin{align}
	u(t,x)&=w(t,x)+\int_0^xl(x,y)w(t,y)dy, \label{eq:inverse_transformation1}\\
	\nonumber v(t,x)&=z(t,x) + \int_0^1 \delta(x,y)w(t,y)dy\\
		&+D\int_0^x p(x,y)z(t,y)dy, \label{eq:inverse_transformation2}
\end{align}
with
\begin{equation}\label{eq:kernel_heat_inverse_bessel}
	l(x,y)=-\lambda y \tfrac{J_1\left( \sqrt{\lambda(x^2-y^2)}\right)}{\sqrt{\lambda(x^2-y^2)}},
\end{equation}
on $\mathcal{T}:=\{(x,y):0\leq y< x\leq 1\}$ where $\mathrm{J}_1(\cdot)$ denotes the  Bessel function of first kind and
\begin{align}
	\label{eq:kernel_delta}	\delta(x,y)&=2 \sum_{n=1}^{\infty} e^{-Dn^2\pi^2 x }\sin(n\pi y)\int_0^1\sin(n\pi \zeta) l(1,\zeta) d \zeta,\\
			p(x,y)&=-\delta_y(x-y,1).
\end{align} Using the estimates of the backstepping transformations, i.e., 
\begin{align}
	\Vert w(t) \Vert_2 &\leq \tilde{k} \Vert u(t) \Vert_2, \label{eq:equivalence_w_u}\\ 
	\Vert u(t) \Vert_2 &\leq \tilde{l} \Vert w(t) \Vert_2, \label{eq:equivalence_u_w}\\
	\Vert z(t) \Vert_{\infty} &\leq \tilde{\gamma} \Vert u(t) \Vert_2 +  \tilde{g}\Vert v(t) \Vert_{\infty}, \label{eq:equivalence_z_v} \\ 
	\Vert v(t) \Vert_{\infty} &\leq \tilde{\delta} \Vert w(t) \Vert_2 +  \tilde{p}\Vert z(t) \Vert_{\infty}, \label{eq:equivalence_v_z}
\end{align}
with $\tilde{k} := 1+ \left(\displaystyle \int_{0}^{1} \left( \int_{0}^{x}\vert k(x,y) \vert^2 dy \right) dx \right)^{1/2}  $,  $\tilde{l} := 1+ \left(\displaystyle\int_{0}^{1} \left( \displaystyle\int_{0}^{x}\vert l(x,y) \vert^2 dy \right) dx \right)^{1/2}$ , $\tilde{g} : =1 + D \displaystyle\max_{0\leq x \leq 1} \int_{0}^{x}\vert g(x,y) \vert dy$,  $\tilde{p}:= 1 + D \displaystyle\max_{0\leq x \leq 1} \int_{0}^{x}\vert p(x,y) \vert dy$,  $\tilde{\gamma}:= \displaystyle\sup_{x \in[0, D]} (\Vert \gamma(x,\cdot) \Vert_2)$, and $\tilde{\delta}:= \displaystyle\sup_{x \in[0, D]}(\Vert \delta(x,\cdot) \Vert_2)$ one establishes the following inequality 
\begin{equation}\label{equivalence}
	M_2\|(u,v)\|\le \|(w,z) \|\le M_1\|(u,v)\|,
\end{equation}
where $M_1,M_2$ are 
\begin{align}
	\label{M1}	M_{1} & =\max\{ (\tilde{k} +\tilde{\gamma}), \tilde{q} \},\quad M_{2}= \tfrac{1}{\max\{ (\tilde{l} +\tilde{\delta}) \}}.
\end{align}
Although \eqref{eq:direct_transformation1}--\eqref{M1} are well-known facts, we present them here as the constants $M_1$ and $M_2$ are incorporated in the control design.
\subsection{Properties of the Quantizer}
The state $u$ of the plant and the actuator state $v$ are available only in quantized form. We consider here dynamic quantizers with an adjustable parameter of the form (see, e.g., \cite{brockett2000quantized,fkoudohode2024,liberzon2003hybrid})
\begin{equation}\label{quantizer}
	q_{\mu}(u,v)=( q_{1\mu}(u),q_{2\mu}(v))=\left(\mu q_1\left(\frac{u}{\mu}\right),\mu q_2\left(\frac{v}{\mu}\right)\right),
\end{equation} where $\mu>0$ can be manipulated and this is called \enquote{zoom} variable. The quantizer $q_1: L^2([0,1]; \mathbb{R}) \to L^2([0,1]; \mathbb{R})$ is Lipschitz on bounded sets (see, e.g. \cite{xu2020}), while $q_2: \mathbb{R} \to \mathbb{R}$ is locally Lipschitz. Both quantizers satisfy the following properties\\
P1: If $\|(u,v)\| \leq M$, then $\|(q_1(u)-u, q_2(v)-v)\| \leq \Delta$,\\
P2: If $\|(u,v)\|>M$, then $\|(q_1(u),q_2(v))\|>M-\Delta$,\\
P3: If $\|(u,v)\| \leq \hat{M}$, then $q_1(u)=0\footnote{\text{The equality $q_1(u) = 0$ in P3 is understood in the sense of distributions.}}$ and $q_2(v)=0,$ \\
for some positive constants $M, \hat{M}$, and $\Delta$, with $M>\Delta$ and $\hat{M}<M$.

Considering the $L^2-$norm of the reaction-diffusion PDE state requires to define the properties of the quantizer accordingly in $L^2$-norm, as well as to impose the additional assumption that the quantizer function also belongs to $L^2$ (that is tacitly assumed in the definition of $q_1$). An example of a quantizer that satisfies this property is when the quantizer may be assumed to also satisfy a specific sector condition (such as, e.g., the logarithmic quantizer \cite{garcia2012model}). In order to actually implement a quantizer using typical quantizer functions (see, e.g., \cite{liberzon2003hybrid}), when the quantizer is defined using an $L^2-$norm, one would have to also use, e.g., a density argument, to guarantee the approximation of its $L^2-$norm by the Euclidean norm of a finite-dimensional counterpart. The work in \cite{wakaiki2024stabilization} is relevant here, as it introduces and implements quantizers in different norms. We also note that we choose to define the properties of the quantizer in terms of the norm of the complete infinite-dimensional state of the PDE-PDE system, as the system considered is equipped with this norm, in correspondence with respective definitions in the finite-dimensional case, see, e.g., \cite{liberzon2003hybrid}.

We further notice that we consider a case in which the measured transport PDE state is also subject to quantization. If the transport PDE state corresponds to a delayed state and if the delayed state is available without quantization, one could employ the past input values in the predictor-feedback law.  Moreover, as in \cite{fkoudohode2024}, we assume a single tunable parameter $\mu$ for simplicity in control design and analysis, which is also practically reasonable in scenarios where, for example, a single computer with a single camera is used for measurements collection.

\subsection{Predictor-Feedback Law Using Quantized Measurements}
The hybrid predictor-feedback law 
is defined as \begin{equation}\label{control_quantizer}
	U(t)=\left\{\begin{array}{ll}0, & 0 \leq t <t_0 \\  P_{\mu(t)}\left(u(\cdot,t),v(\cdot,t)\right), & t\ge t_0
	\end{array}\right.,
\end{equation} where
\begin{align}\label{predictor_quantizer}
\nonumber	P_{\mu}(u,v)&=\int_0^1\gamma (1,y)q_{1\mu}(u(y))dy\\
	&+D\int_0^{1}g(1,y)q_{2\mu}(v(y))dy.
\end{align} 
The tunable parameter $\mu$ is selected as 
\begin{equation}\label{switching_parameter}
	\mu(t)= \begin{cases}\overline{M}_1 \mathrm{e}^{2\sigma_1 (j+1) \tau} \mu_{0}, & (j-1) \tau \leq t < j \tau+\bar{\tau}\delta_j, \\ & 1 \leq j \leq\left\lfloor\frac{t_0}{\tau}\right\rfloor,\\ \mu\left(t_0\right), & t \in[t_0, t_0+T), \\ \Omega \mu \left(t_0+(i-2) T\right), & t \in [t_0+(i-1) T, \\ & t_0+i T), \quad i=2,3, \ldots\end{cases},
\end{equation}   for some fixed, yet arbitrary, $\tau, \mu_0>0$, \textcolor{black}{where $t_0=m\tau+\bar{\tau},$ for an $m\in\mathbb{Z}_+,$ $\bar{\tau}\in[0,\tau),$ and $\delta_m=1,\delta_j=0,j< m$}, with $t_0$ being the first time instant at which the following holds\footnote{Note that \eqref{first_time_t0} can be verified using only the availabe quantized measurements.} 
\begin{align}
	\nonumber&\left\|\mu(t_0)q_1\left(\frac{u\left(t_{0}\right)}{\mu\left(t_{0}\right)}\right)\right\|_2+\left\|\mu(t_0)q_2\left(\frac{v\left( t_{0}\right)}{\mu\left(t_{0}\right)}\right)\right\|_{\infty}\\
	&	\label{first_time_t0}\leq (M \overline{M}- \Delta) \mu(t_0),
\end{align} where
\begin{align}
	\label{M3}	M_{3} & =\|\gamma(1,\cdot) \|_2+D\max_{0 \leqslant y \leqslant 1}|g(1,y)|  , \\
	\label{M}	\overline{M} & =\tfrac{M_{2}}{M_1(1+M_0)}, \\
	\label{Mbar} \overline{M}_1&=\max\{\sqrt{2},G+1\},\\
	\label{C}	G\footnotemark &=4 \sqrt{\sum_{n=0}^{\infty}\tfrac{n^2\pi^2}{(\lambda-n^2\pi^2)^{2}}},\\
	\label{A}\sigma_1&=\lambda-\pi^2,\\
	\label{Omega}	\Omega & =\frac{(1+\lambda_1) (1+M_0)^2 \Delta M_{3}}{M_{2} M}, \\
	\label{T}	T & =-\frac{\ln \left(\frac{\Omega}{1+M_0} \right)}{\delta}.
\end{align}
\footnotetext{For simplicity of derivations within the proof of Lemma~\ref{Lemma1} it is tacitly assumed that $\lambda\neq n^2\pi^2$ for all $n$. If it happens that $\lambda=\pi^2\bar{n}^2$ for some $\bar{n}$, one could show (with additional, but tedious computations) that \eqref{normX0t0} still holds with $\bar{M}_1 =\max\{\sqrt{2},G+1\} $ and $G =4 \sqrt{\displaystyle\sum_{n\neq \bar{n}}\tfrac{n^2\pi^2}{(\lambda-n^2\pi^2)^{2}}}+2\sqrt{2}\pi \bar{n}.$}
The parameters $\delta, \lambda_1$ and $M_0$ are defined as follows.
Parameter $\delta\in (0, \min\{\pi^2,\nu\})$, $\lambda_1$ is selected large enough in such a way that the following small-gain condition holds
\begin{equation}\label{small_gain}
	\frac{1}{1+\lambda_1}<\frac{e^{-D}}{1+\frac{\sqrt{3}}{3}},	
\end{equation} and $M_0$ is defined such that
\begin{align}
	\nonumber& M_{0}= (1-\varphi_1)^{-1}\max \left\{1; \tfrac{1}{\sqrt{3}}(1+\varepsilon)(1-\phi)^{-1}e^{D(\nu+1)}\right\}\\
	&+\left(1-\phi\right)^{-1}\left(1-\varphi_1\right)^{-1} \max\left\{e^{D(\nu+1)} ;\phi \right\},
\end{align} where $0<\phi<1$ and $0<\varphi_{1}<1$ with
\begin{align} \label{phi and psy}
	\phi=\frac{1+\varepsilon}{1+\lambda_1} e^{D(\nu+1)} \text{ and }\varphi_{1}=\tfrac{1}{\sqrt{3}}(1+\varepsilon)(1-\phi)^{-1}\phi,
\end{align} for some  $\varepsilon, \nu>0$.  The choice of $\nu,\varepsilon$ is such that it guarantees that $\phi<1,\varphi_1<1$, which is always possible given \eqref{small_gain}. 

\section{Stability of Switched Predictor-Feedback Controller Under State Quantization}\label{stabStateQuantization}
\begin{theorem}\label{Theorem1}
	Consider the closed-loop system consisting of the plant \eqref{eq:sysparabolic1_cascade}--\eqref{eq:BC_hyperbolic_cascade} and the switched predictor-feedback law \eqref{control_quantizer}--\eqref{switching_parameter}.  If $\Delta$ and $M$ satisfy  \begin{equation}\label{conditionMDelta}
		\frac{\Delta}{M}<\frac{M_2}{(1+M_0)\max \{M_3(1+\lambda_1)(1+M_0),2M_1\}},
	\end{equation}	then for every initial data $v_0 \in C_{\rm rpw}([0, 1]; \mathbb{R})$  and  $u_0 \in L^2(0, 1)$,
 the solution $(u,v)$  to \eqref{eq:sysparabolic1_cascade}--\eqref{eq:BC_hyperbolic_cascade} satisfies the property
	\begin{align}
		\nonumber& \Vert u(t) \Vert_2 + \Vert v(t)\Vert_{\infty} \\
		\label{stability_result_u} 	&\leq  \gamma\left( \Vert u_0 \Vert_2 + \Vert v_0\Vert_{\infty} \right)^{\left(2-\frac{\ln \Omega}{T} \frac{1}{\sigma_1}\right)} \mathrm{e}^{\frac{\ln \Omega}{T}t},
	\end{align}	where
	\begin{align}
		\nonumber	\gamma&=\frac{\overline{M}_1}{M_2}\max \left\{\frac{M_{2}M}{\Omega} e^{2\sigma_1 \tau} \mu_{0}, M_1\right\} \\
		\nonumber&
		\times \max \left\{\frac{1}{\mu_0(M \overline{M}-2 \Delta)}, 1\right\}\\
		&\times\left(\frac{1}{\mu_0(M \overline{M}-2 \Delta)}\right)^{\left(1-\frac{\ln \Omega}{T} \frac{1}{\sigma_1}\right)}.
	\end{align} 
%
\end{theorem}
Although we do not state an existence and uniqueness result, as its proof is out of the scope of the present paper, which focuses on the control design and stability analysis, in principle, it can be studied as follows. Within the interval $[0, t_0)$, explicit solutions (thanks to spectral analysis and the method of characteristics), together with \cite[Corollary 2.2]{karafyllis2018sampled}, can be used to show that for every $v_0 \in C_{\rm rpw}( [0,1])$ and $u_0 \in L^2(0,1)$, the solution to the open-loop system satisfies $u \in C\left([0,t_0);L^2(0,1)\right)$ and $v  \in C_{rpw}([0,t_0)\times [0, 1]))  $. 

The well-posedness of the system \eqref{eq:sysparabolic1_cascade}--\eqref{eq:BC_hyperbolic_cascade} and \eqref{control_quantizer} for $t \ge  t_0$ can be proved by induction, starting in the interval $[t_0, t_0 + T)$ (where $\mu$ is constant)  as key step. Thanks to the Lipschitzness of the mapping $\varphi_\mu: L^2([0, 1] ; \R) \times L^{\infty}([0, 1] ; \R)\to \R$ defined as $\varphi_\mu(z,w)=\int_0^1\gamma (1,y)\mu q_1\left(\frac{z(y)}{\mu}\right)dy+D\int_0^{1}g(1,y)\mu q_2\left(\frac{w(y)}{\mu}\right)dy$ (given the Lipschitzness assumptions on $q_1$ and $q_2$), and similarly to the proof of \cite[Theorem 11.3]{karafyllis2021input} we can apply Banach's fixed-point theorem (constructing proper mappings based on explicit solutions to \eqref{eq:sysparabolic1_cascade}--\eqref{eq:BC_hyperbolic_cascade}) to obtain a unique (local) solution to \eqref{eq:sysparabolic1_cascade}--\eqref{eq:BC_hyperbolic_cascade} and \eqref{control_quantizer} satisfying $u \in C\left([t_0, t_0 + T_1);L^2(0,1)\right) $ and $v \in \mathcal{C}_{\text{rpw}}([t_0, t_0 + T_1)\times [0,1])$ for some $0<T_1<T$. The (global) well-posedness on $[t_0,t_0+T)$ follows from the boundedness of the solution $(u, v)$ in $L^2\times L^{\infty}$ (thanks to Lemma \ref{Lemma1} and Lemma \ref{Lemma2}). By induction, we obtain existence and uniqueness of a solution $u \in C\left(\mathbb{R}_+;L^2(0,1)\right) $ and $v \in \mathcal{C}_{\text{rpw}}(\mathbb{R}_+\times [0,1])$.

The following lemmas provide critical bounds on both the open-loop and closed-loop phases of the system, which are essential for establishing stability of the solutions.

\blem\label{Lemma1}
	Let $\Delta$ and $M$ satisfy \eqref{conditionMDelta}, 
	there exists a time $t_{0}$ satisfying 
	\begin{equation}\label{t0}
		t_{0} \leqslant \frac{1}{\sigma_1} \ln\left(\dfrac{\frac{1}{\mu_{0}}\left(\Vert u_0 \Vert_2 + \Vert v_0\Vert_{\infty} \right)}{(M \overline{M}-2 \Delta)}\right),
	\end{equation} 
	such that \eqref{first_time_t0} holds, and thus, the following also holds 
	\begin{equation}\label{bound_in_time_t0}
		\Vert u(t_{0}) \Vert_2 + \Vert v(t_{0})\Vert_{\infty}  \leq  M \overline{M}\mu(t_{0}).
	\end{equation}
\elem
\begin{proof}	
	For all $0\le t< t_{0}$, thanks to \eqref{control_quantizer} one has $U(t)=0,$ and thus, the corresponding open-loop system reads as
	\begin{eqnarray}
		u_t(t,x) & =&   u_{xx}(t,x) + \lambda u(t,x), \label{eq:sysparabolic1_cascade_openLoop}\\
		u(t,0)&=&0, \label{eq:BC_parabolic1_cascade_openLoop0} \\
		u(t,1)&=& v(t,0),  \label{eq:BC_parabolic1_cascade_openLoop1} \\
		v_{t}(t,x)&=& \tfrac{1}{D}v_{x}(t,x),\label{eq:hyperbolic_cascade_openLoop}\\
		v(t,1)&=&0.\label{eq:BC_hyperbolic_cascade_openLoop}
	\end{eqnarray} 
	The solution to the transport subsystem \eqref{eq:hyperbolic_cascade_openLoop}, \eqref{eq:BC_hyperbolic_cascade_openLoop}, using the method of characteristics is given as $v(t,x) = v_0(\tfrac{1}{D}t+x)$ for $ t \leq D(1-x)$ and  $v(t,x)= 0$ for $t \geq D(1-x)$.  
	Therefore, 
	\begin{equation}\label{estimation_u_norm}
		\|v(t)\|_{\infty} \leqslant \left\|v_{0}\right\|_{\infty}. 
	\end{equation}
			From the equations \eqref{eq:sysparabolic1_cascade_openLoop}--\eqref{eq:BC_parabolic1_cascade_openLoop1} we use \cite[Identity (3.11)]{karafyllis2016ISS} for $0\le t< t_{0}$, $\left\|u(t)\right\|_2=\sqrt{\sum_{n=1}^{\infty}\left|c_{n}(t)\right|^{2}} $	with $c_n(t)=e^{\sigma_n t} c_n(0)-\frac{d \phi_n(1)}{d x} \displaystyle\int_0^t e^{\sigma_n(t-s)} v(s,0)d s$ and 
			\begin{align}
				\begin{split}
					|c_n(t)|&\leq e^{\sigma_n t}\left|c_{n}(0)\right|+\left|\frac{d}{d x} \phi_n(1)\right| \frac{|1-e^{\sigma_n t}|}{\sigma_{n}}\\
					&\times \operatorname{ess\ sup}_{0 \leqslant s \leqslant t}|v(s,0)|,
				\end{split}
			\end{align} 
		where $\sigma_{n}=\lambda-\pi^2n^2,\quad \phi_n(x)=\sqrt{2}\sin(\pi n x), n=1,2,\dots.$ 
			Applying the Young's inequality and using the inequalities
		$e^{\sigma_n t}=e^{\lambda t}e^{-\pi^2 n^2 t}\leq e^{\lambda t}e^{-\pi^2 t}=e^{\sigma_1t}$  and $\left|1-e^{\sigma_n t}\right|\le 1+e^{\sigma_n t}\le 1+e^{\sigma_1 t}\le 2 e^{\sigma_1 t}$ we obtain  		
		\begin{align}
			\begin{split}
				\left|c_{n}(t)\right|^{2} &\leqslant 2 e^{2\sigma_1 t}\left|c_{n}(0)\right|^{2}+\frac{8}{\sigma_{n}^{2}}\left|n\pi\sqrt{2}\cos(\pi n)\right|^2\\&\times e^{2\sigma_1 t} \left(\operatorname{ess\ sup}_{0 \leqslant s \leqslant t}\left|v(s,0)\right|\right)^2
			\end{split}
		\end{align} Therefore,
\begin{equation}
			\left\|u(t)\right\|_{L^{2}} \leq \sqrt{2} e^{\sigma_1 t}\| u_{0} \|_{2}+Ge^{\sigma_1 t} \left\|v_{0}\right\|_{\infty}, \label{normXt0}
	\end{equation} where  $ G$ and $\sigma_1 $ are given by \eqref{Mbar} and \eqref{C} respectively.
		Therefore, combining \eqref{estimation_u_norm} and \eqref{normXt0}  we have for $0\le t< t_{0}$
		\begin{equation}\label{normX0t0}
			\Vert u(t) \Vert_2 + \Vert v(t)\Vert_{\infty}   \leqslant \overline{M}_1e^{\sigma_1 t}\left(\Vert u_0 \Vert_2 + \Vert v_0\Vert_{\infty}  \right),
		\end{equation} with $\overline{M}_1$ defined in \eqref{A}.
Choosing the switching signal $\mu$ according to \eqref{switching_parameter}, one has the existence of a time $t_{0}$ verifying \eqref{t0}  such that
	\begin{align} 
		\dfrac{\left\|u\left(t_{0}\right)\right\|_2}{\mu(t_{0})}+	 \dfrac{\left\|v\left(t_{0}\right)\right\|_{\infty}}{\mu(t_{0})} \le M \overline{M}-2 \Delta.
	\end{align} The rest of the proof is identical to that of Lemma 5 in \cite{fkoudohode2024IMA}.
\end{proof}
\blem\label{Lemma2}
	Select $\lambda_1$ large enough in such a way that the small-gain condition \eqref{small_gain} holds. Then the solutions to the target system \eqref{eq:Target_sysparabolic1_cascade}--\eqref{eq:Initial_cond_target_z} with the quantized controller \eqref{control_quantizer}, which verify, for fixed $\mu$,
	\begin{equation}\label{hyplemma2}
		\|w(t_{0})\|_2+\|z(t_{0})\|_{\infty}\le \frac{M_{2}}{1+M_0} M\mu,
	\end{equation} they satisfy for $t_{0}\le t< t_{0}+T$
	\begin{align}
		\nonumber&	\|w(t)\|_2+\|z(t)\|_{\infty}\leqslant \max\left\{ M_{0} e^{-\delta (t-t_{0})}\left(\left\|w(t_{0})\right\|_2 \right.\right.\\
		&\left. +\left\|z(t_{0})\right\|_{\infty}\right),\left. \Omega \frac{M_{2}}{1+M_0} M\mu \right\}.\label{normwz}
	\end{align} In particular, the following holds
	\begin{align}\label{normXu1}
		\|w(t_{0}+T)\|_2+\|z(t_{0}+T)\|_{\infty} \leq \Omega \frac{M_{2}}{1+M_0} M\mu.
	\end{align}
\elem
\begin{proof}
	See the proof of Lemma 6 in \cite{fkoudohode2024IMA}.
\end{proof}

{\em Proof of Theorem~\ref{Theorem1}:} See the proof of Theorem 3  in \cite{fkoudohode2024IMA}.
{\hfill $\Box$ } 
%
%
\section{Extension to Input Quantization}\label{inputquantization}
The results in the previous section could be extended to input quantization by modifying the switched predictor-feedback law as follows
\begin{equation}\label{control_quantizerinput}
	U(t)=\left\{\begin{array}{ll}0, & 0 \leq t < \bar{t}_{0} \\  \bar{q}_{\mu}\left(U_{\rm nom}(t)\right), & t\ge\bar{t}_{0}
	\end{array}\right.,
\end{equation} where $U_{\rm nom}(t)$ is given in \eqref{nominalU} and the quantizer is a locally Lipschitz function $\bar{q}_{\mu}:\mathbb{R}\to \mathbb{R},$ defined by $\bar{q}_{\mu}(\bar{U})=\mu \bar{q}\left(\tfrac{\bar{U}}{\mu}\right)$, satisfying the following properties\\
$\bar{\rm P}1$: If $|\bar{U}| \leq M$, then $|\bar{q}(\bar{U})-\bar{U}| \leq \Delta$,\\
$\bar{\rm P}2$: If $|\bar{U}|>M$, then $|\bar{q}(\bar{U})|>M-\Delta$,\\
$\bar{\rm P}3$: If $|\bar{U}| \leq \hat{M}$, then $\bar{q}(\bar{U})=0$.

Considering the parameters defined in \eqref{M3}--\eqref{T}, the switching variable $\mu$ is selected as 
\begin{equation}\label{switching_parameterinput}
	\mu(t)= \begin{cases}\overline{M}_1 \mathrm{e}^{2\sigma_1 (j+1) \tau} \mu_{0}, & (j-1) \tau \leq t < j \tau+\bar{\tau}_1\delta_j, \\ & 1 \leq j \leq\left\lfloor\frac{\bar{t}_0}{\tau}\right\rfloor,\\ \mu\left(\bar{t}_0\right), & t \in[\bar{t}_0, \bar{t}_0+T), \\ \Omega \mu \left(\bar{t}_0+(i-2) T\right), & t \in [\bar{t}_0+(i-1) T, \\ & \bar{t}_0+i T), \quad i=2,3, \ldots\end{cases},
\end{equation} 
for some fixed, yet arbitrary, $\tau, \mu_0>0$, \textcolor{black}{ where $\bar{t}_0=\bar{m}\tau+\bar{\tau}_1,$ for an $\bar{m}\in\mathbb{Z}_+,$ $\bar{\tau}_1\in[0,\tau),$ and $\delta_{\bar{m}}=1,\delta_j=0,j< \bar{m}$}, with $\bar{t}_0$ being the first time instant at which the following event is detected using the available measurements of the actuators states are available, holds  
%
\begin{equation}\label{bound_in_time_t0input}
	\|u(\bar{t}_{0})\|_2+\|v(\bar{t}_0)\|_{\infty}\leq  \dfrac{M \overline{M}}{M_3}\mu(\bar{t}_0).
\end{equation}	
\begin{theorem}\label{Theorem2}
	Consider the closed-loop system consisting of the plant \eqref{eq:sysparabolic1_cascade}--\eqref{eq:BC_hyperbolic_cascade} and the switched predictor-feedback law \eqref{control_quantizerinput},    \eqref{switching_parameterinput} with \eqref{nominalU}.  If $\Delta$ and $M$ satisfy  \begin{equation}\label{conditionMDeltainput}
		\frac{\Delta}{M}<\frac{M_2}{M_3(1+\lambda_1)(1+M_0)^2},
	\end{equation}	
	then for every initial data $v_0 \in C_{\rm rpw}([0, 1]; \mathbb{R})$ and  $u_0 \in L^2(0, 1)$ the solution  $(u,v)$ to \eqref{eq:sysparabolic1_cascade}--\eqref{eq:BC_hyperbolic_cascade} satisfies 
\begin{align}
	\nonumber	& \Vert u(t) \Vert_2 + \Vert v(t)\Vert_{\infty}
	\label{stability_result_q} \\
	&\leq  \bar{\gamma}\left( \Vert u_0 \Vert_2 + \Vert v_0\Vert_{\infty} \right)^{\left(2-\frac{\ln \Omega}{T} \frac{1}{\sigma_1}\right)} \mathrm{e}^{\frac{\ln \Omega}{T}t},
\end{align}	where
\begin{align}
	\nonumber	\bar{\gamma} &=\tfrac{\sigma_1}{M_2}\max \left\{\tfrac{M_{2}M}{\Omega M_3} e^{2\sigma_1 \tau} \mu_{0}, M_1\right\} \max \left\{\tfrac{M_3}{\mu_0M \overline{M}}, 1\right\}\\
	&
	\times \left(\tfrac{M_3}{\mu_0M \overline{M}}\right)^{\left(1-\tfrac{\ln \Omega}{T} \tfrac{1}{\sigma_1}\right)}.
\end{align} 	
\end{theorem}	
To prove Theorem~\ref{Theorem2}, we first establish two lemmas, whose proofs employ reasoning similar to that used in the case of state quantization.
\blem\label{Lemma3}
There exists a time $\bar{t}_0$ satisfying 
\begin{equation}\label{t0input}
	\bar{t}_{0} \leqslant \frac{1}{\sigma_1} \ln\left(\tfrac{\frac{M_3}{\mu_{0}}\left(\left\|u_{0}\right\|_2+\left\|v_{0}\right\|_{\infty}\right)}{M \overline{M}}\right),
\end{equation} 
such that \eqref{bound_in_time_t0input} holds.		
\elem
\begin{proof}
For all $0 \leq t <\bar{t}_0$, the system is described by \eqref{eq:sysparabolic1_cascade_openLoop}--\eqref{eq:BC_hyperbolic_cascade_openLoop}.Thus, we obtain, as in the proof of Lemma \ref{Lemma1}, estimate \eqref{normX0t0}. By selecting the switching signal $\mu$ according to \eqref{switching_parameterinput}, there exists a time $\bar{t}_0$ that satisfies \eqref{t0input}, ensuring that the relation \eqref{bound_in_time_t0input} holds.
\end{proof}
\blem\label{Lemma4}
Select $\lambda_1$ large enough in such a way that the small-gain condition \eqref{small_gain} holds. Then, the solutions to the target system \eqref{eq:Target_sysparabolic1_cascade}--\eqref{eq:Target_BC_hyperbolic_cascade} with the quantized controller \eqref{nominalU}, \eqref{control_quantizerinput}, \eqref{switching_parameterinput}, which verify, for fixed $\mu$,
\begin{equation}\label{hyplemma4input}
	\|w(\bar{t}_{0})\|_2+\|z(\bar{t}_{0})\|_{\infty}\le \frac{M_{2}M\mu}{(1+M_0)M_3},
\end{equation} they satisfy for $\bar{t}_0\le t<\bar{t}_0+T$
\begin{align}
	\nonumber& \|w(t)\|_2+\|z(t)\|_{\infty}\leqslant \max\left\{ M_{0} e^{-\delta (t-\bar{t}_{0})}\left(\left\|w(\bar{t}_{0})\right\|_2 \right.\right.\\
	&\left. +\left\|z(\bar{t}_{0})\right\|_{\infty}\right),\left.  \frac{\Omega M_{2}M\mu }{(1+M_0)M_3}  \right\}.\label{norm_X_winput}
\end{align} In particular, the following holds
\begin{align}\label{normXu1input}
	\|w(\bar{t}_{0}+T)\|_2+\|z(\bar{t}_{0}+T)\|_{\infty} \leq  \frac{\Omega M_{2}M\mu}{(1+M_0)M_3}.
\end{align}
\elem

\begin{proof}
 Using the same strategy, as in the proof of  Lemma~\ref{Lemma2} the following inequalities hold
\begin{equation}\label{norm_x_winput}
	\Vert w \Vert_{[\bar{t}_{0},t]} \le \|w(\bar{t}_{0})\|_2+ \tfrac{1}{\sqrt{3}}(1+\varepsilon)\Vert z \Vert_{[\bar{t}_{0},t]}, 
\end{equation}
\begin{equation}
	\begin{split}
		\Vert z \Vert_{[\bar{t}_0,t]} \leq & e^{D\left(\nu+1\right)} \Vert z(\bar{t}_0) \Vert_{\infty}+e^{D\left(\nu+1\right)}\\&\times(1+\varepsilon) \operatorname{ess\sup}_{\bar{t}_0 \leqslant s \leqslant t}\left(\left|\bar{d}(s)\right| e^{\delta (s-\bar{t}_0)}\right), \label{normzd}
	\end{split}
\end{equation}
where for $U_{\rm nom}$ and $\mu$ given in \eqref{nominalU} and \eqref{switching_parameterinput}, respectively,
\begin{align}
	\bar{d}(t)=\mu(t)\bar{q}\left(\dfrac{U_{\rm nom}(t)}{\mu(t)}\right)-U_{\rm nom}(t).	\label{d1input}
\end{align} 
	Provided that 
	\begin{align}
			\label{conditioninput}
			\frac{\Omega}{(1+M_0)^2}\frac{M_{2}}{M_3} M\mu \leq\|w\|_2+\|z\|_{\infty} \leqslant \frac{M_{2}}{M_3} M\mu,
		\end{align}
	using the property $\rm \bar{P}1$ of the quantizer, the left-hand side of bound \eqref{equivalence}, and the definition \eqref{Omega}, we obtain from \eqref{d1input} \begin{align}
			\nonumber	\left|\bar{d}\right|
			& \leqslant \frac{1}{1+\lambda_1}\left(\|w\|_2+\|z\|_{\infty}\right). 
\end{align}	
Repeating the respective arguments from the proof of Lemma~\ref{Lemma2}, we arrive at
\begin{equation}\label{norm_X_u1input}
	\|w(t)\|_2+\|z(t)\|_{\infty}\leqslant M_0e^{-\delta (t-\bar{t}_0)}\left(|X(\bar{t}_0)|+\left\|w(\bar{t}_0)\right\|_{\infty}\right). 
\end{equation}
Thus, for $\bar{t}_0\le t < \bar{t}_0 +T $, using relation \eqref{hyplemma4input} we get
\begin{align}
	\|w(t)\|_2+\|z(t)\|_{\infty}
	\label{forMaxTerminput}	& \le \frac{M_2}{M_3}M\mu, 
\end{align} which makes estimate \eqref{norm_X_u1input} legitimate. Moreover, thanks to relation \eqref{hyplemma4input} and \eqref{T}, one obtains from \eqref{norm_X_u1input} \begin{align}
	\left\|w\left(\bar{t}_0+T\right)\right\|_2+\left\|z\left(\bar{t}_0+T\right)\right\|_{\infty} \label{norm_X_u_Tinput}&\leqslant  \frac{\Omega M_2M\mu}{(1+M_0)M_3}, 
\end{align}  and hence, bound \eqref{normXu1input} is obtained.
The rest of the proof utilizes the same reasoning as the proof of Lemma~\ref{Lemma2}.
\end{proof}

{\em Proof of Theorem~\ref{Theorem2}:} The proof of Theorem~\ref{Theorem2} follows a similar approach to that used in the corresponding part of the proof of Theorem~\ref{Theorem1}, with Lemmas~\ref{Lemma3} and~\ref{Lemma4} playing analogous roles to Lemmas~\ref{Lemma1} and~\ref{Lemma2}.

\section{Conclusions and Future Work}\label{conlusion}
In this paper, we achieved the global asymptotic stability of an unstable reaction-diffusion PDE with input delay under both state and input quantization, developing a switched predictor-feedback control law and combining the backstepping method with small-gain and input-to-state stability techniques.  
Future research could explore designing event-triggered controllers under quantization, inspired by \cite{koudohode2024}. The challenges for such a design would be the construction of the event-triggering mechanism such that it depends only on quantized measurements, as well as the proof of Zeno phenomenon avoidance, due to the non-differentiability of the quantizer function.

	\bibliography{RD-Quantizer}
\end{document}